\documentclass[twocolumn]{aastex62}

\usepackage{multirow}
\usepackage{amsmath}
\usepackage{longtable}

\newcommand{\twco} {^{12}\mbox{CO}}
\newcommand{\twcoa} {^{12}\mbox{CO}~(1-0)}
\newcommand{\twcob} {^{12}\mbox{CO}~(2-1)}
\newcommand{\thco} {^{13}\mbox{CO}}
\newcommand{\ceio} {\mbox{C}^{18}\mbox{O}}

\newcommand{\h} {^{\mathrm{h}}}
\newcommand{\m} {^{\mathrm{m}}}
\newcommand{\s} {^{\mathrm{s}}}

\newcommand{\vlsr}     {v_\mathrm{lsr}}
\newcommand{\vout}     {v_\mathrm{out}}
\newcommand{\tex}     {T_\mathrm{ex}}

\newcommand{\mJybeam}  {\mbox{mJy}~\mbox{beam}^{-1}}

\newcommand{\mJybeamkms}  {\mbox{mJy}~\mbox{beam}^{-1}~\mbox{km s}^{-1}}

\newcommand{\kms}	{\mbox{km s}^{-1}}
\newcommand{\pc}	{\mbox{pc}}
\newcommand{\yr}	{{\rm yr}}

\newcommand{\K}	{{\rm K}}
\newcommand{\au} {\mbox{au}}

\shorttitle{HH 46/47 Molecular Outflow}
\shortauthors{Zhang et al.}

\begin{document}

\title{An Episodic Wide-angle Outflow in HH 46/47}

\author{Yichen Zhang}
\affiliation{Star and Planet Formation Laboratory, RIKEN Cluster for Pioneering Research, Wako, Saitama 351-0198, Japan; yichen.zhang@riken.jp}

\author{H\'ector G. Arce}
\affiliation{Astronomy Department, Yale University, P.O. Box 208101, New Haven, CT 06520, USA}

\author{Diego Mardones}
\affiliation{Departamento de Astronom\'ia, Universidad de Chile, Casilla 36-D, Santiago, Chile}
\affiliation{Centre for Astrochemical Studies, Max-Planck-Institute for Extraterrestrial Physics, Giessenbachstrasse 1, 85748, Garching, Germany}

\author{Sylvie Cabrit}
\affiliation{LERMA, Observatoire de Paris, UMR 8112 du CNRS, ENS, UPMC, UCP, 61 Av. de l'Observatoire, F-75014 Paris, France}
\affiliation{Institut de Plan\'etologie et d'Astrophysique de Grenoble (IPAG) UMR 5274, Grenoble, 38041, France}

\author{Michael M. Dunham}
\affiliation{Department of Physics, State University of New York at Fredonia, 280 Central Avenue, Fredonia, NY 14063, USA}

\author{Guido Garay}
\affiliation{Departamento de Astronom\'ia, Universidad de Chile, Casilla 36-D, Santiago, Chile}

\author{Alberto Noriega-Crespo}
\affiliation{Space Telescope Science Institute, 3700 San Martin Dr., Baltimore, MD 21218, USA}

\author{Stella S. R. Offner}
\affiliation{Department of Astronomy, The University of Texas at Austin, Austin, TX 78712, USA}

\author{Alejandro C. Raga}
\affiliation{Instituto de Ciencias Nucleares, UNAM, Ap. 70-543, 04510 D.F., Mexico}

\author{Stuartt A. Corder}
\affiliation{Joint ALMA Observatory, Av. Alonso de C\'ordova 3107, Vitacura, Santiago, Chile}

\begin{abstract}
During star formation, the accretion disk drives fast MHD winds which usually contain two components, 
a collimated jet and a radially distributed wide-angle wind. 
These winds entrain the surrounding ambient gas producing 
molecular outflows. We report recent observation of $\twcob$ emission of the HH 46/47 molecular outflow by
the Atacama Large Millimeter/sub-millimeter Array,
in which we identify multiple wide-angle outflowing shell structures in both the blue and red-shifted outflow lobes. 
These shells are highly coherent in  position-position-velocity space, extending to $\gtrsim 40-50~\kms$
in velocity and $10^4~\au$ in space with well defined morphology and kinematics.
We suggest these outflowing shells are the result of the entrainment of ambient gas 
by a series of outbursts from an intermittent wide-angle wind. 
Episodic outbursts in collimated jets are commonly observed, 
yet detection of a similar behavior in wide-angle winds has been elusive.
Here we show clear evidence that the wide-angle component of the HH 46/47 protostellar outflows 
experiences similar variability seen in the collimated component. 
\end{abstract}

\keywords{ISM: clouds, Herbig-Haro objects, individual objects (HH 46, HH 47), jets and outflows 
--- stars: formation}

\section{Introduction}
\label{sec:intro}

Outflows play an important role in star formation and the evolution of molecular clouds and cores,
as they remove angular momentum from the accretion disk (e.g. \citealt[]{Bjerkeli16,Hirota17,Lee17,Zhang18}), 
carve out cavities in their parent cores (e.g. \citealt[]{Arce06}),
and inject energy and momentum into the star-forming environments (e.g. \citealt[]{Arce10,Plunkett13}).
During star formation, the accreting circumstellar disk drives bipolar magneto-centrifugal winds (e.g. \citealt[]{Shu00,Konigl00}).
Models predict that these protostellar winds have both collimated
and wide-angle components (e.g., \citealt[]{Kwan95,Shang98,Matt03}).
The collimated portion of the wind, which is usually refereed to as a jet, 
is typically traced by optical line emission in later-stage exposed sources (e.g. \citealt[]{Reipurth01}),
or sometimes in molecular emissions in early-stage embedded sources (e.g. \citealt[]{Tafalla10,Lee17}).
The wide-angle component 
(presumably arising from a larger stellocentric radius in the disk)
 is thought to be slower than the collimated 
component, and does not produce the striking features seen in jets. 
In young embedded sources, the wide-angle component of a disk wind may be detected with high-resolution 
molecular line observations (e.g. \citealt[]{Tabone17,Louvet18}). 
In more evolved pre-main sequence stars this component has been observed with 
optical atomic emission lines (e.g. \citealt[]{Bacciotti00}).

Both jets and wide-angle winds can interact with the ambient molecular gas and 
entrain material to form slower, but much more massive outflows,
which are typically observed in CO and other molecules and are generally referred to as molecular outflows.
The entrainment process is not yet fully understood.
Models include entrainment through jet bow-shocks (internal
and/or leading) (e.g. \citealt[]{Raga93}) and wide-angle winds (e.g. \citealt[]{Li96}).
In the jet bow-shock entrainment model, a jet propagates into the surrounding cloud,  and
forms bow-shocks which push and accelerate the ambient gas producing outflow shells surrounding the jet
(e.g. \citealt[]{Tafalla16}).
In the wide-angle wind entrainment model, a radial wind blows into the ambient material, forming
an expanding outflow shell (e.g. \citealt[]{Lee00}). 
These two mechanisms may act simultaneously since jet and wide-angle wind may co-exist.

The accretion of material from a circumstellar disk 
onto a forming star is believed to be episodic (e.g. \citealt[]{Dunham12}).
The variation in the accretion rate may arise from various instabilities in the accretion disk
(e.g. \citealt[]{Zhu10,Vorobyov15}).
In protostars, which are embedded in their parent gas cores (i.e. the so-called Class 0 and Class I sources),
the most significant evidence of episodic accretion comes from jets that show a series of knots 
(which sometimes are evenly spaced) along their axes (e.g. \citealt[]{Lee07,Plunkett15}).
These knots often trace bow-shocks which are formed by variations in the mass loss rate 
or jet velocity, which in turn may be caused by variation in the accretion rate.
However, such variability has not yet been seen
in  wide-angle outflows, which in principle
should experience the same variations as jets.

Here we report  recent $\twcob$ observations of the HH 46/47 molecular outflow using
the Atacama Large Millimeter/sub-millimeter Array (ALMA) that reveal
multiple wide-angle outflowing shells, which we argue were formed by an
episodic wide-angle wind.
The HH 46/47 outflow is driven by a low-mass early Class I source 
(HH 47 IRS, which is also known as HH 46 IRS 1, IRAS 08242-5050)
with a bolometric luminosity of $L_\mathrm{bol}=12~L_\odot$ that resides in the 
Bok globule ESO 216-6A, located on the outskirts of the 
Gum Nebula at a distance of 450 pc (\citealt[]{Schwartz77,Reipurth00,Noriega04}).
Previous ALMA $\twcoa$ observations (\citealt[]{Arce13,Zhang16}, referred to as Papers I and II hereafter) 
showed a highly asymmetric CO outflow
(with the red-shifted lobe extending a factor of four more than the blue-shifted lobe), 
as the driving source lies very close to the edge of the globule.
In addition to the wide molecular outflow, 
collimated jets are also optically seen in the blue-shifted lobe 
(\citealt[]{Reipurth91,Eisloffel94,Hartigan05})
and in the infrared in the red-shifted lobe 
(\citealt[]{Micono98,Noriega04,Velusamy07}).
Detailed analysis of the morphology and kinematics of the molecular outflow showed evidence
of wide-angle wind entrainment for the blue-shifted outflow lobe and jet bow-shock entrainment for the 
red-shifted lobe (see Papers I and II).
The difference between the two molecular outflow lobes is likely due to the fact that the blue-shifted jet 
is mostly outside of the globule where the outflow cavity has little or no molecular gas inside, 
while the red-shifted jet is pushing through the core, surrounded by dense gas. 
However, even in the red-shifted side, the energy distribution shows that more energy is injected by the outflow
at the base of the outflow cavity, which is consistent with a wide-angle wind entrainment scenario,
rather than at the jet bow-shock heads, as a jet-entrainment scenario would suggest.

\section{Observations}
\label{sec:obs}

\begin{figure*}
\begin{center}
\includegraphics[width=0.7\textwidth]{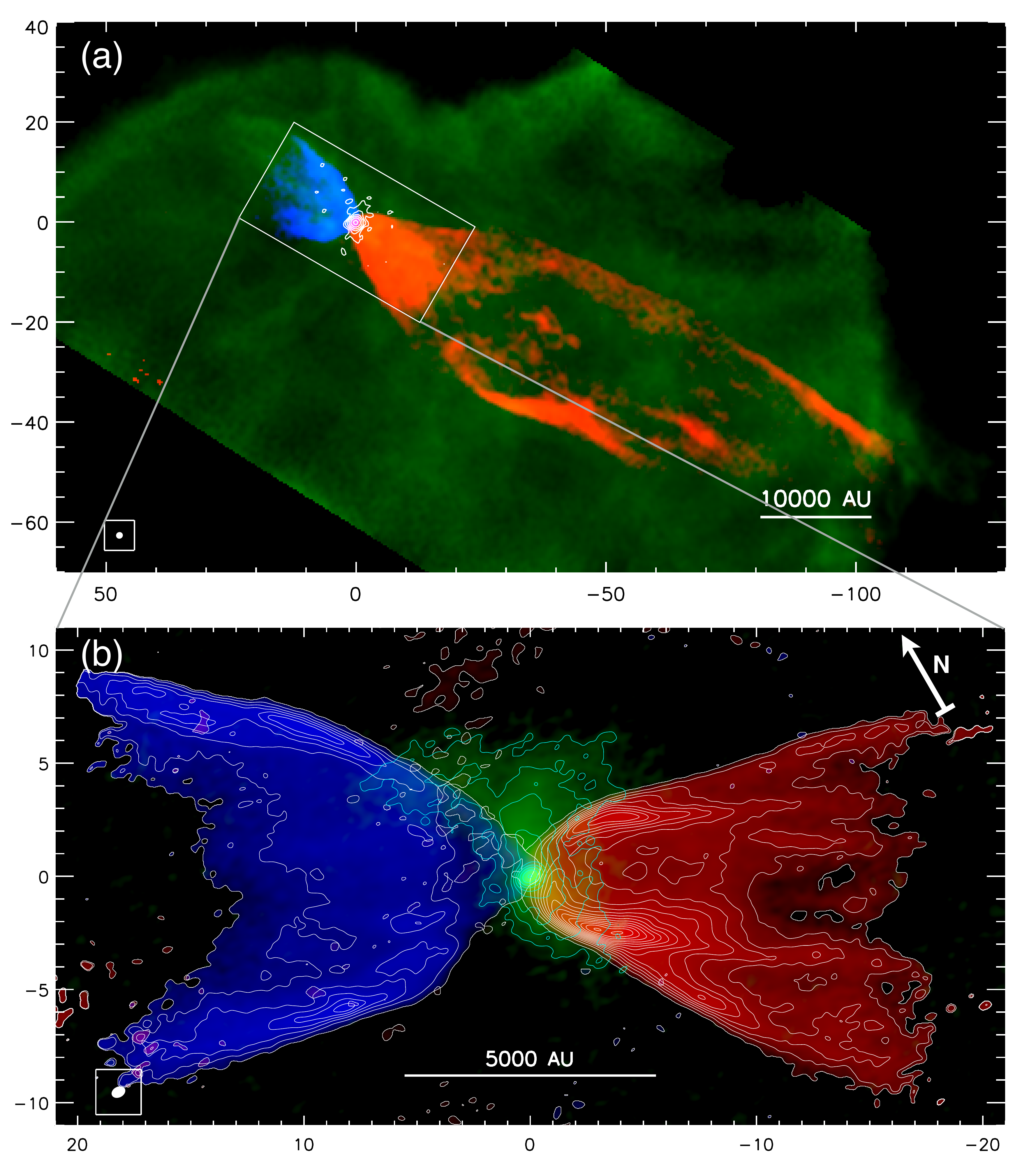}\\
\caption{
{\bf (a):} $\twcoa$ integrated intensity map of the HH 46/47 molecular outflow at large scales from 
Figure 1 of Paper II.
The red, blue, and green color scales show emission integrated over the velocity ranges from 1 to 10 
$\kms$, from $-10$ to $-1~\kms$, and from $-0.6$ to $0.6~\kms$ (relative to the cloud velocity), respectively.
The synthesized beam is $1.37\arcsec\times1.31\arcsec$.
The white contours show the 100 GHz continuum emission.
{\bf (b):} $\twcob$ 
integrated intensity map of the
HH 46/47 molecular outflow overlaid on the 1.3mm dust continuum emission. 
The blue-shifted lobe is integrated from $\vout=-35$ to $-10~\kms$, 
and the red-shifted lobe is integrated
from $\vout=+10$ to $+50~\kms$. 
The contours start at $5\sigma$ and have intervals of $30\sigma$ 
($1\sigma=4.3~\mJybeamkms$ for the blue-shifted lobe and $1\sigma=4.9~\mJybeamkms$
for the red-shifted lobe),
The green color shows the 1.3 mm continuum emission with contour levels of 
$(2^n)\times 5\sigma$ ($n=0, 1, 2, 3, ..., 8$) with $1\sigma=0.021~\mJybeam$.
The image are rotated by 30$^\circ$ counterclockwise.
The synthesized beam of the $\twco$ map is $0.67\arcsec\times0.48\arcsec$, and is shown in the lower-left corner of the panel.}
\label{fig:figure1}
\end{center}
\end{figure*}

The observations were carried out using ALMA band 6 on Jan 6, 2016 with the C36-2 configuration
and on June 21, 30 and July 6, 2016 with the C36-4 configuration 
(as part of observations for project 2015.1.01068.S).
In the C36-2 configuration observation, 36 antennas were used and 
the baselines ranged from 15 to 310 m. The total on source integration time was 75 min.
J1107-4449 and J0538-4405 were used as bandpass and flux calibrators, and
J0811-4929 and J0904-5735 were used as phase calibrator.
In the C36-4 configuration observations, 36 antennas were used and
the baselines ranged from 15 to 704 m. The total integration time was 150 min.
J1107-4449, J0538-4405, and J1107-4449 were used as bandpass and flux calibrators, and
J0811-4929 was used as phase calibrator. 
The observations included only one pointing
centered at $8\h25\m43\s.8$, $-51^\circ00\arcmin36\arcsec.0$ (J2000) 
which is the 3mm continuum peak obtained from the Cycle 1 observations (Paper II).
The primary beam size (half power beam width) is about 23$\arcsec$ at Band 6.

The $\twcob$ emission at 230.54 GHz was observed
with a velocity resolution of about 0.09 $\kms$. 
The center of the $\twco$ spectral window, which has a bandwidth of 
117 MHz ($\sim 150~  \kms$), is shifted from the $\twcob$ line central 
frequency by 18 MHz ($\sim 23~\kms$) in order to
observe both $\twco$ and $\thco$ lines in one spectral setup. 
As a result, our $\twco$ observation covers emission 
from $\vlsr= -94$ to $+56~\kms$.
The $\thco$ (2-1), $\ceio$ (2-1), H$_2$CO ($3_{0,3}-2_{0,2}$), and CH$_3$OH ($4_{2,2}-3_{1,2}$) lines
were observed simultaneously in the same spectral set-up.
In addition, a spectral window with a bandwidth of 1875 MHz was used to map the 1.3 mm continuum. 
In this paper we focus on the $\twcob$ and continuum data.
We defer the discussions of other molecular lines to a future paper.

The data were calibrated and imaged in CASA (\citealt[]{McMullin07}; version 4.5.3). 
Self-calibration was applied by using the continuum data 
after normal calibration.
The task CLEAN was used to image the data. For the spectral data we defined
a different clean region for each channel. 
Robust weighting with a robust parameter of 0.5 was used in the CLEAN process.
The resulting synthesized beam is $0.65\arcsec \times 0.47\arcsec$ (P.A. = $87.8^\circ$) for the continuum data,
and $0.67\arcsec \times 0.48\arcsec$ (P.A. = $88.2^\circ$) for the $\twco$ data.
Throughout the paper we define the outflow velocity $\vout$ as the LSR velocity of the emission minus
the cloud LSR velocity, which is 5.3 $\kms$ (\citealt[]{vanKempen09}).

\section{Results}
\label{sec:result}

\begin{figure*}
\begin{center}
\includegraphics[width=0.9\textwidth]{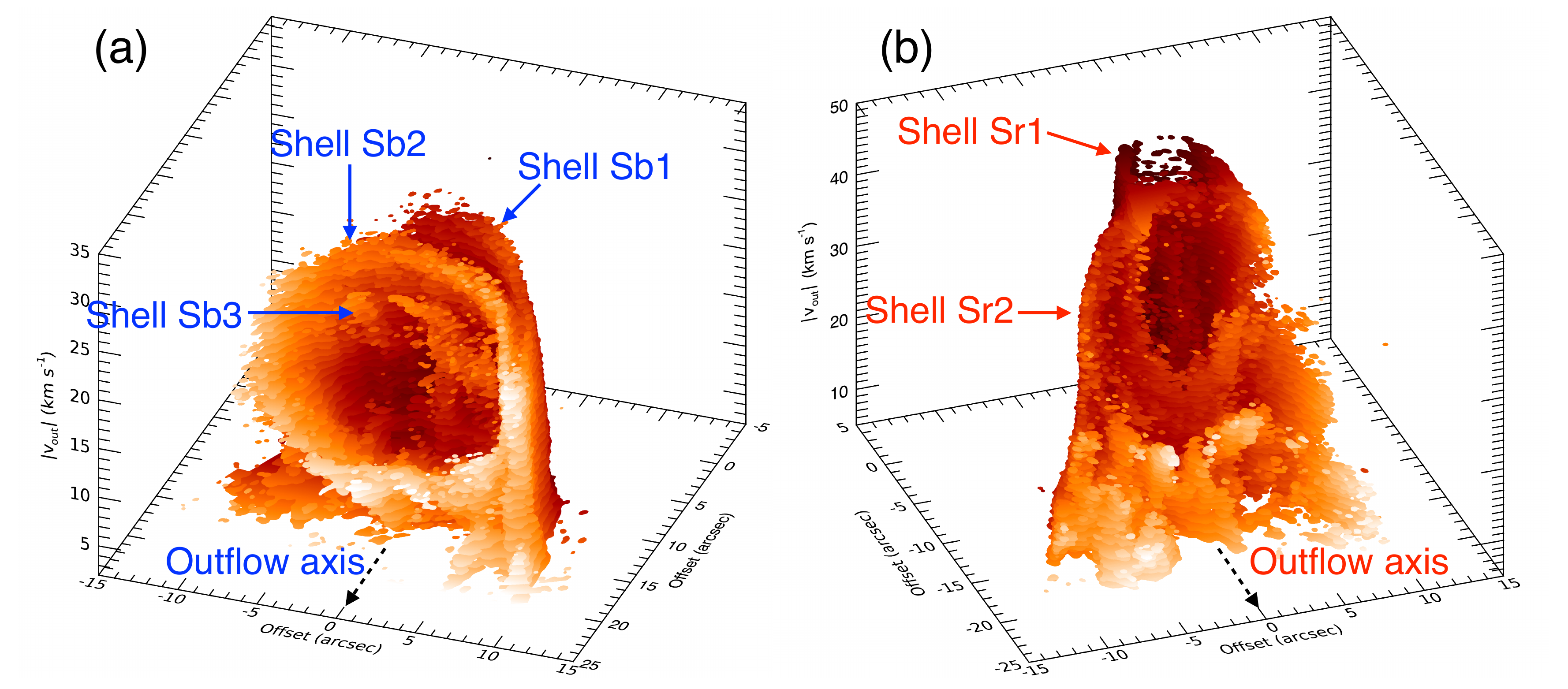}\\
\caption{
{\bf (a):}  Position-position-velocity diagram for the $\twcob$ blue-shifted lobe. 
Channels at $|\vout|>2~\kms$ with emission higher than $5\sigma$ 
($1\sigma=1.6~\mJybeam$ for a channel width of $0.3~\kms$) are included.
The emission outside of the outflow cavity is not included. 
The position of the outflow source, at offset (0,0), is at the back, and the outflow direction is toward the reader.
The color is selected to emphasize the layered structure.
{\bf (b):} Same as panel (a) but for the red-shifted lobe. 
In this panel, data with velocities $|\vout|>5~\kms$ are included.}
\label{fig:PPV}
\end{center}
\end{figure*}

\begin{figure*}
\begin{center}
\includegraphics[width=\textwidth]{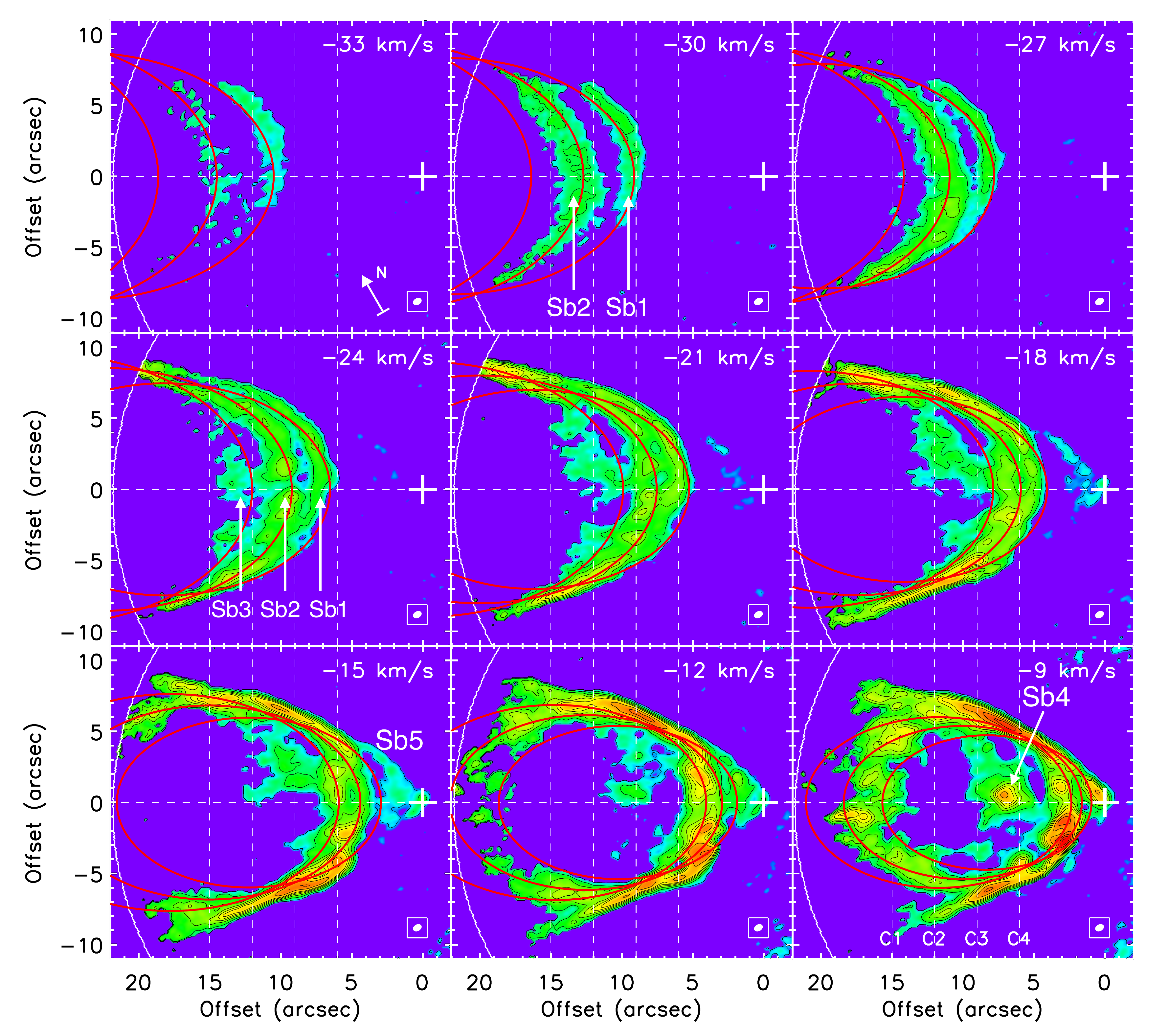}\\
\caption{Channel maps of the $\twcob$ emission of the blue-shifted lobe.
The central velocity of each 3~$\kms$-wide channel is indicated in upper right corner of each  panel. 
The maps are rotated by $30^\circ$ so that the outflow axis is along the $x$-axis.
The contours are at levels of 5$\sigma$, 20$\sigma$, 40$\sigma$, 60$\sigma$, 80$\sigma$, 100$\sigma$, 150$\sigma$, 200$\sigma$,
250$\sigma$, 300$\sigma$, and 350$\sigma$ with $1\sigma=0.52~\mJybeam$.
The white crosses mark the position of the 1.3 mm continuum peak. 
The white curve to the left of each panel indicates the edge of the field of view with a primary beam gain greater than 0.1.
The horizontal and vertical dash lines show the cuts for the position-velocity diagrams in Figure \ref{fig:pvdiagram}.
The major shell structures are labeled.
Red ellipses are the ``best-fit'' models for shells Sb1, Sb2 and Sb3.}
\label{fig:chanmap_B}
\end{center}
\end{figure*}

\begin{figure*}
\begin{center}
\includegraphics[width=\textwidth]{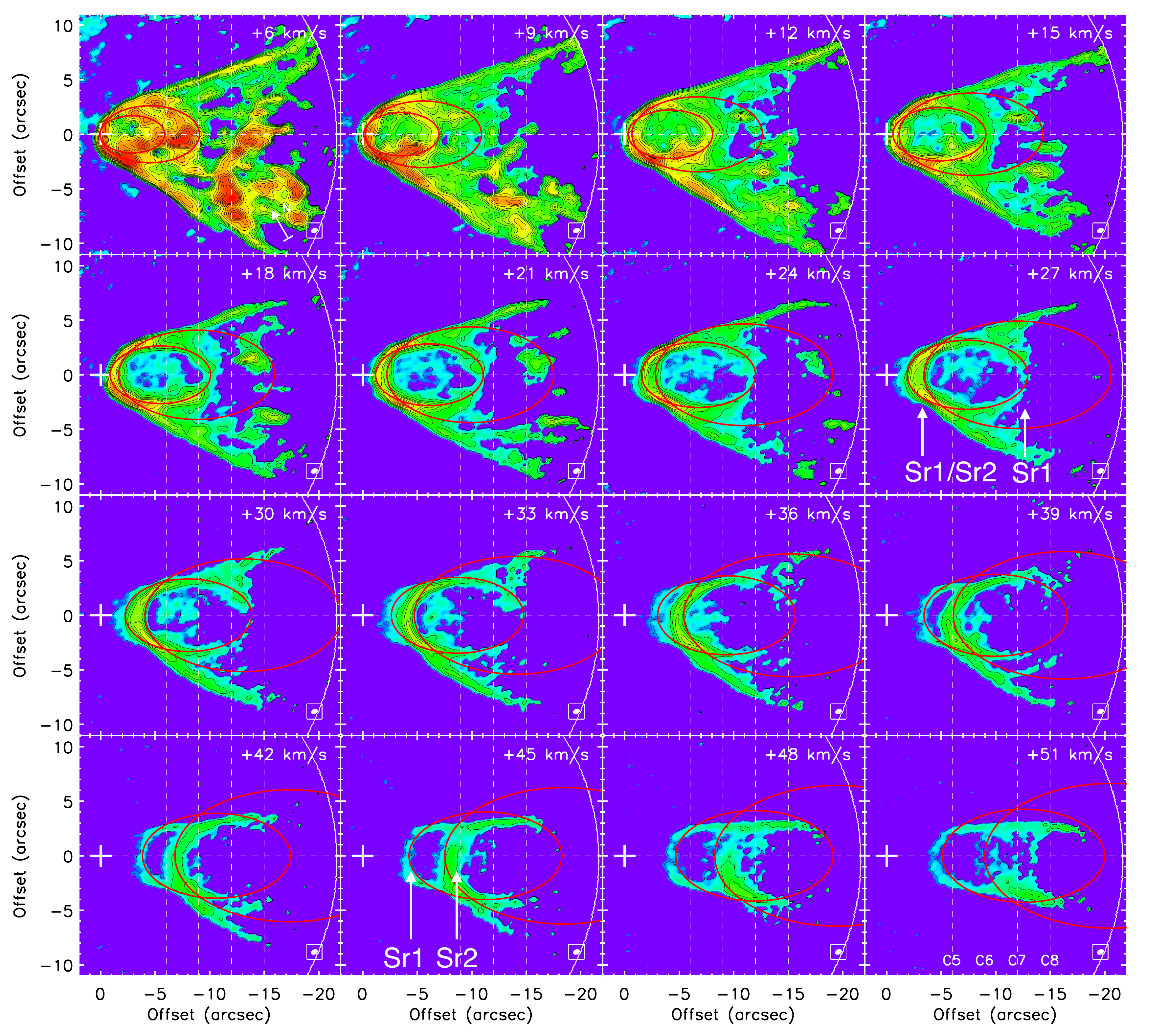}\\
\caption{Same as Figure \ref{fig:chanmap_B}, but for the red-shifted lobe.}
\label{fig:chanmap_R}
\end{center}
\end{figure*}

\begin{figure*}
\begin{center}
\includegraphics[width=0.77\textwidth]{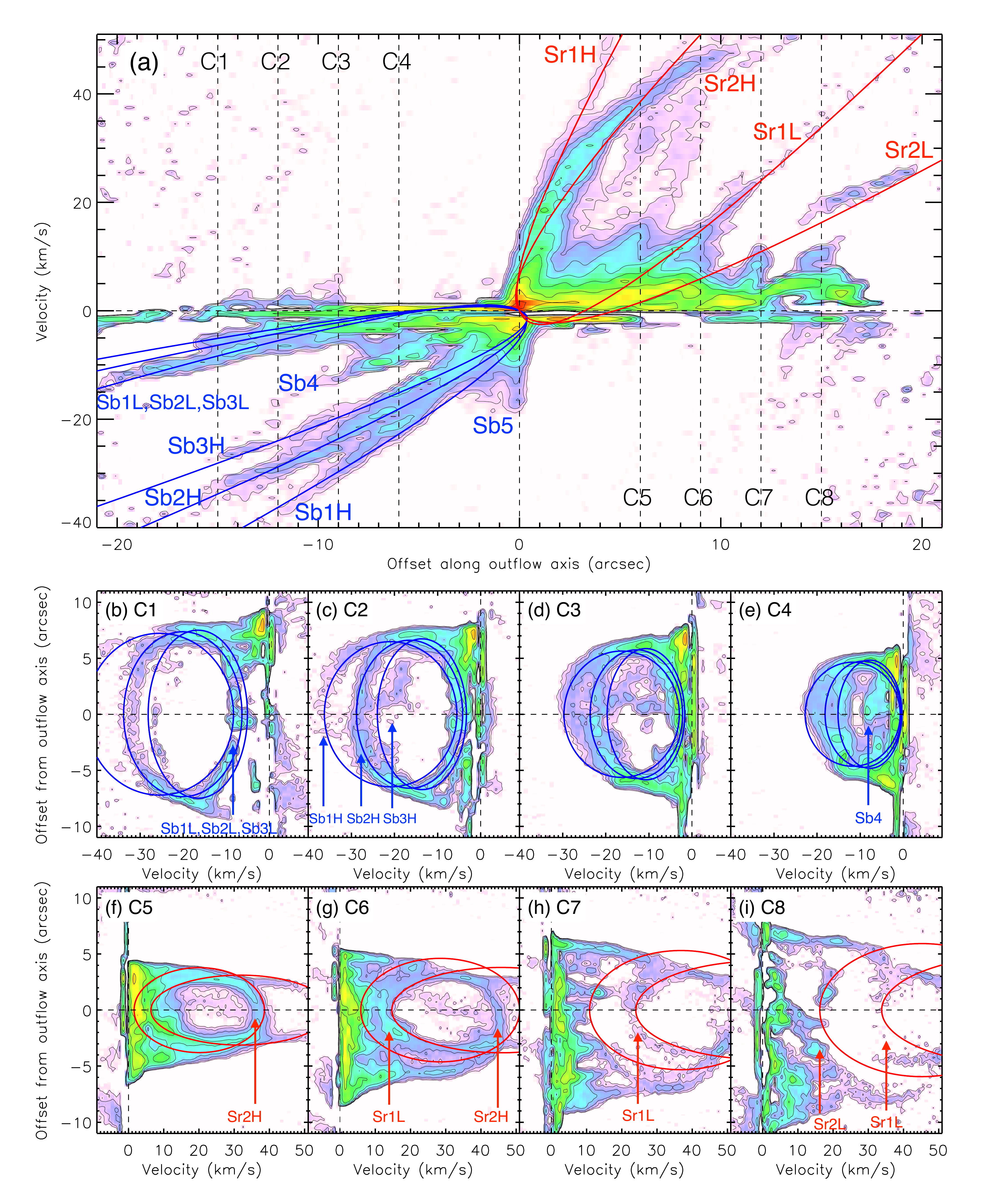}\\
\caption{{\bf (a):} Position-velocity diagram of the $\twcob$ emission along the outflow axis with a cut width of 1$\arcsec$.
{\bf (b)-(i):} Position-velocity diagrams of the $\twcob$ emission along 1$\arcsec$-wide cuts perpendicular to the outflow axis.
The cuts are shown in Figure \ref{fig:chanmap_B} and \ref{fig:chanmap_R}.
The offset positions of the cuts perpendicular to the outflow axis are also shown in panel (a).
The contours are at levels of 
$(2^n)\times 5\sigma$ ($n=0, 1, 2, 3, ...$) with 1$\sigma=0.56~\mJybeam$.
The major structures are labelled. The red and blue curves are the ``best-fit'' models for shells Sb1, Sb2 and Sb3 (blue lines) and shells Sr1 and Sr2 (red lines).}
\label{fig:pvdiagram}
\end{center}
\end{figure*}

Figure \ref{fig:figure1} shows the integrated intensity maps of the 
blue-shifted and red-shifted $\twcob$ emission.
Unlike the previous $\twcoa$ observations (shown in panel a), our $\twcob$ single ALMA pointing observations 
only allow us to detect the outflow emission up to about $20\arcsec$ away from the protostar. Both lobes show 
conical morphologies of similar size,  in contrast to what is seen when the full extent of the two lobes is observed. 
The $\twcob$ emission is also
more symmetric with respect to the outflow axis
than the $J=1-0$ emission in which the northern side of the blue-shifted outflow
is much brighter than its southern side.
In Figure \ref{fig:figure1}(b), the red lobe appears to be composed of different shell structures. 
At a distance of $7\arcsec$ from the central source the inner shells delineate a U-like structure with a width of 
about $6\arcsec$ inside a cone-like shell that is about $9\arcsec$ wide. Although multiple shells are not
clearly seen in the integrated image of the blue lobe, they are seen in the channel maps 
(see Figure \ref{fig:chanmap_B}). 
The 1.3 mm continuum emission shows an elongated structure perpendicular to the outflow axis,
which is consistent with previous observation of the 3 mm continuum. 
The extended continuum emission appears to be shaped by the 
outflow, as it approximately follows the shape of the outflow cavity. 

The shell structures are best seen in the position-position-velocity (PPV) diagrams (Figure \ref{fig:PPV}),
where they appear to be highly coherent in space and velocity.
At least two shells can be identified in each lobe: Sb1 and Sb2 in the blue-shifted lobe (panel a); 
and Sr1 and Sr2 in the red-shifted lobe (panel b).  
There is possibly a third shell in the blue-shifted outflow (Sb3), 
that appears to have a more complex structure (i.e., less coherent structure) than the other shells.
Each of these shells shows a cone-like shape in the PPV space (best seen in shell Sr1 ans Sb2), 
with a high-velocity side and a low-velocity side
(also see Figure \ref{fig:pvdiagram}).
In the red-shifted lobe, both high-velocity and low-velocity sides of Sr1 and Sr2 are distinguishable 
(Figure \ref{fig:PPV}b).
However, in the blue-shifted lobe, while the high-velocity sides of Sb1 and Sb2 are clearly separate,
the low-velocity sides of the these two shells seem to have merged.
We expect the  high-velocity side of a shell seen in PPV space to correspond to 
the front side of the blue-shifted shell or the back side of the red-shifted shell 
as the expanding motion of the outflow shell, in addition to the outflowing motion, 
is contributing to the observed line-of-sight velocities.
On the other hand, at a particular line-of-sight velocity these shells have shapes similar to
ellipses, partial ellipses, or parabolas (see Figures \ref{fig:chanmap_B} and \ref{fig:chanmap_R}).
Therefore, structures seen in different positions and velocities
can come from a single coherent structure.

In addition to the velocity field within one shell, the overall velocities of the shells are different
from each other, which is shown by their different opening directions in the PPV space.
For example, in the red-shifted lobe, shell Sr1 is generally faster than Sr2
(i.e. the velocity of the Sr1 shell at any distance from the protostar
is higher than that of the Sr2 shell at the same distance), while in the blue-shifted lobe,
shell Sb1 is generally faster than Sb2 (see also Figure \ref{fig:pvdiagram}). 
The shape of the shells is similar, but some have different widths.  In the red-shifted lobe, 
shell Sr1 is much narrower than shell Sr2 
(see also Figure \ref{fig:chanmap_R}). Since shell Sr1 is faster and narrower than Sr2, 
the two shells intersect in  PPV space (Figure \ref{fig:PPV}a).
In the blue-shifted lobe, however, the shells appear to have similar widths.
At low velocities (in the lower part of the two PPV diagrams in Figure \ref{fig:PPV}), 
the emission becomes complex and has many sub-structures, 
therefore it cannot be clearly identified as being part of one of the shells identified at higher velocities. 

In Figures \ref{fig:chanmap_B} and \ref{fig:chanmap_R} we plot
the channel maps of the $\twcob$ emission.
Significant emission in the blue-shifted lobe is detected up to about $\vout=-35~\kms$, 
even though the spectral window covers velocities up to $\vout=-99~\kms$.
In this outflow lobe, the emission moves away from the central source as the velocity increases.
At blue-shifted outflow velocities of about $\vout=-35~\kms$ the emission is found at the edge of our map. 
Thus it is probable that there exists higher velocity outflow emission beyond the border of our map. 
In the red-shifted lobe, the emission is still quite strong at the edge of the spectral window,
which only covers up to outflow velocities of about $\vout=+51~\kms$.
Hence, we suspect the red-shifted lobe extends to even higher velocities.  

The shell structures identified in the PPV diagrams are clearly seen
in these channel maps, and are labeled in the figures. 
In the blue-shifted outflow, shells Sb1 and Sb2
can be easily distinguished at velocities $\vout=-33$ to $-20~\kms$.
Shell Sb3 is seen at velocities $\vout=-24$ to $-21~\kms$.
At $\vout=-18$ to $-9~\kms$, the emission inside the cavity
is actually from a structure different from Sb3 (best seen in Figure \ref{fig:pvdiagram}), 
which we label Sb4.
At these relatively lower outflow velocities, the shells Sb1, Sb2 and Sb3 appear to merge together 
and show a full elliptical shape.
It is not clear whether the far side of the ellipse corresponds to the low-velocity side 
of one of the Sb1, Sb2 or Sb3 shells, 
or a structure produced from the combination of these three shells (also see Figure \ref{fig:pvdiagram}).
At velocities of $|\vout|<20~\kms$, 
additional emission appears close the central source (which we label Sb5), 
showing a cone shape rather than an elliptical shape. 
The major structures Sb1, Sb2 and Sb3, 
all shift to the northeast (i.e., left in Figure~\ref{fig:chanmap_B}) and become wider as the 
outflow velocity increases.

In the red-shifted outflow, the two main shell structures Sr1 and Sr2 
are best seen in the outflow velocity range from $+30$ to $+50~\kms$.
As discussed above, the widths of the two main shells are
quite different. 
Although in general, the bulk of the emission shifts away from the source as the outflow velocity increases,
the position of the narrower shell (Sr1), does not change much.
As discussed above, the two shells intersect in the PPV space and this is most clearly 
seen in the $\vout=39$ to $45~\kms$
channels in Figure \ref{fig:chanmap_R}.
At  low velocities (e.g. $|\vout|<15~\kms$), the Sr1 shell can still be discerned even though
a significant amount of material fills the outflow cavity.

Figure \ref{fig:pvdiagram} shows the position-velocity (PV) diagrams along the outflow axis
and perpendicular to the outflow axis. 
They correspond to the intersections of the PPV diagram with different position-velocity planes.
As discussed above, a shell in the PPV space has a high velocity side and a low velocity side, which
becomes evident in the PV diagrams. We label different structures in  
Figure \ref{fig:pvdiagram} with ``H'' or ``L'' to indicate the high and low velocity sides of the same shell.
In the red-shifted lobe,
pairs of high and low velocity structures of the same shell are easily identified (Sr1H/Sr1L and Sr2H/Sr2L).
 There is also emission between the Sr2H and Sr1L structures, 
and emission at larger distances with low velocities,
which cannot be identified as part of a shell.
In the blue-shifted lobe, while the high-velocity sides of the Sb1, Sb2 and Sb3 shells are easily distinguished,
their corresponding low-velocity walls are not so clear.
It is unclear whether the structure labeled as ``Sb1L, Sb2L, Sb3L'' 
corresponds to the low-velocity side of one of the three shells 
(Sb1, Sb2 or Sb3)
or if this structure is produced by the merger of the low-velocity side of all three shells. 
It is also unclear whether Sb4 and Sb5 are separate structures or the low and high velocity  sides of  
a single shell.

\section{Discussion}
\label{sec:discussion}

\subsection{Shell Model Fitting}
\label{sec:fitting}

\begin{table*} 
\scriptsize
\begin{center}
\caption{Parameters of the fitted parabolic shells \label{tab:model}}
\begin{tabular}{ccccccc}
\hline
\hline
Shell & $i$ $^\mathrm{a}$ & $R_0$ (arcsec) & 
$t_0$ ($\mathrm{arcsec~km}^{-1}~\mathrm{s}$) & age ($10^3$ years)  $^\mathrm{b}$ 
& $v_0$ ($\kms$) $^\mathrm{c}$ &
$\theta_\mathrm{open}$ ($z=15\arcsec$) $^\mathrm{d}$ \\
\hline
Sb1 & \multirow{3}{*}{$40^\circ$} & 2.6 & 0.55 & 1.2 & 4.7 & 22.6$^\circ$ \\
Sb2 & & 2.7 & 0.70 & 1.5 & 3.9 & 23.0$^\circ$ \\
Sb3 & & 2.8 & 0.85 & 1.8 & 3.3 & 23.3$^\circ$ \\
\hline
Sr1 & \multirow{2}{*}{$35^\circ$} & 1.3 & 0.15 & 0.32 & 8.7 & 16.4$^\circ$ \\
Sr2 & & 1.9 & 0.25 & 0.53 & 7.6 & 19.6$^\circ$ \\
\hline
\end{tabular}
\end{center}
$^\mathrm{a}$  Inclination angle between the outflow axis and the plane of sky. 
The same value is used for shells in the same lobe in order to reduced the numbers of free parameters.\\
$^\mathrm{b}$  Dynamical age calculated from $t_0$ assuming a distance of $d=450~\pc$.\\
$^\mathrm{c}$  Characteristic velocity of the shell defined as $v_0\equiv R_0/t_0$.\\
$^\mathrm{d}$  Half opening angle of the fitted shell at a height of $z=15\arcsec$, 
$\tan\theta_\mathrm{open}=\sqrt{R_0/15\arcsec}$.
\end{table*}

To be more quantitative, we fit the morphology and kinematics
of the outflow shells with expanding parabolas. 
Following the method by \citet[]{Lee00}, 
the morphology and velocity of a single expanding parabolic shell 
can be described (in cylindrical coordinates) as
\begin{equation}
\left(\frac{z}{R_0}\right)=\left(\frac{R}{R_0}\right)^2, \quad v_z=\frac{z}{t_0}, \quad v_R=\frac{R}{t_0},
\end{equation}
where the $z$-axis is along the outflow axis,
the $R$-axis is perpendicular to the $z$-axis,
and $v_z$ and $v_R$ are the velocities in the directions of $z$ and $R$
(i.e., the forward velocity and the expansion velocity), respectively.
The free parameters in this  model are the inclination $i$ between the outflow axis
and the plane of the sky, parameter $R_0$ which determines the width of the outflow shell, 
and $t_0$ which determines the velocity distribution of the outflow shell.
Note that the characteristic radius $R_0$ is 
just the radius of the shell at $z=R$, i.e. $\theta=45^\circ$.
Also $t_0$ can be considered as the dynamical age of the shell.
We further define a characteristic velocity $v_0\equiv R_0/t_0$,
which is the outflowing velocity $v_z$ or expanding velocity $v_R$ at $z=R$.
In such a model,  the half opening angle of the shell at the height of $z$ is
$\tan\theta_\mathrm{open}(z)=\sqrt{R_0/z}$.

Such a model predicts an elliptical shape of emissions in the channel maps,
and as the channel velocity increases the elliptical structure becomes wider and shifts further
away from the central source.
The model also predicts a parabolic shape in the PV diagram along the outflow axis
and elliptical shapes in the PV diagrams perpendicular to the outflow axis.
Such behaviors are indeed consistent with our observations.
The same model was used to explain the blue-shifted lobe in Paper II,
in which we did not have enough spatial resolution and sensitivity to detect
the multiple shell structures.

We fit the shells Sb1 and Sb2 in the blue-shifted lobe
and shells Sr1 and Sr2 in the red-shifted lobe 
by comparing the model described above with the observed emission distributions
in both channel maps and PV diagrams.
Here we only focus on the location of the emission in  space and velocity, and
do not attempt to reproduce the intensity distribution. 
To reduce the number of free parameters, we adopt a constant
inclination angle for  shells in the same lobe.
To perform the fitting, the inclination $i$ is searched within a range 
$30^\circ\leq i \leq 45^\circ$ with an interval of $5^\circ$,
the parameter $t_0$ is searched within a range 
from 0.05 to 1 $\mathrm{arcsec~km}^{-1}~\mathrm{s}$
with an interval of $0.05~\mathrm{arcsec~km}^{-1}~\mathrm{s}$,
and the parameter $R_0$  is searched with an interval of 0.01$\arcsec$ within a range
from $1\arcsec$ to $3\arcsec$.
Furthermore, in fitting the shells Sb1 and Sb2, 
we assume that the low-velocity walls of these two shells
have merged into one structure (labeled as ``Sb1L, Sb2L, Sb3L'' in Figure \ref{fig:pvdiagram}).

The best-fit models are selected by visually comparing the model curves (which
are shown by the red curves in Figures \ref{fig:chanmap_B} and \ref{fig:chanmap_R}, 
and the blue and red curves in Figure \ref{fig:pvdiagram})
with the observed distribution of the outflow emission.
The parameter values of what we consider the ``best-fit'' models are listed in Table \ref{tab:model},
including the characteristic velocities $v_0=R_0/t_0$ and 
the shell half opening angles $\theta_\mathrm{open}$ 
at $z=15\arcsec$.
The fitted inclinations are $i=40^\circ$ and $i=35^\circ$ between the outflow axis
and the plane of sky for the blue-shifted and red-shifted outflows, respectively.
These are consistent with the values derived from observations
of the optical (blue-shifted) jet by \citet[]{Eisloffel94} and \citet[]{Hartigan05}, which are $34^\circ\pm3^\circ$ and
$37^\circ.5\pm2^\circ.5$, respectively.

In the red-shifted outflow, the Sr1 and Sr2 shells are fit with
$t_0=0.15$ and $0.25~\mathrm{arcsec~km}^{-1}~\mathrm{s}$ and
$R_0=1.3\arcsec$ and $1.9\arcsec$.
These models describe the two shells relatively well, especially the Sr1 shell.
The model fit to Sr2 is not as good as that of Sr1, especially at higher velocities. 
This is partly due to the fact that Sr2 is slightly asymmetric with respect to the outflow axis 
(seen more clearly in  the channel maps at $\vout=+33~\kms$ to $+45~\kms$ Fig.~\ref{fig:chanmap_R}).
Sr2 also appears tilted (or skewed) in the PV diagrams perpendicular to the outflow axis 
(panels h and i of Figure \ref{fig:pvdiagram}). These features may be caused by rotation 
or a slight change in the outflow direction, which the models do not take into account. 
The fitted values of $t_0$ of 0.15 and $0.25~\mathrm{arcsec~km}^{-1}~\mathrm{s}$
correspond to time scales of $3.2\times 10^2$ and $5.3\times 10^2~\mathrm{yr}$
assuming a source distance of 450 pc,
which can be considered as the dynamical ages of these two shells,
result in an age difference between the Sr1 and Sr2 shells of 
$2.1\times 10^2~\mathrm{yr}$.

In the blue-shifted side, the parameters for the best-fit models for shells Sb1 and Sb2 are
$t_0=0.55$ and $0.70~\mathrm{arcsec~km}^{-1}~\mathrm{s}$ and
$R_0=2.6\arcsec$ and $2.7\arcsec$.
The fitted values of $t_0$ correspond to time scales of $1.2\times 10^3$ and $1.5\times 10^3~\mathrm{yr}$,
which result in an age difference between the Sb1 and Sb2 shells of 
$3.2\times 10^2~\mathrm{yr}$.
If we assume that the time interval between shells Sb1 and Sb2 is the same as the interval between 
Sb2 and Sb3, then we can estimate a value for $t_0$ for shell Sb3 of 
$0.85~\mathrm{arcsec~km}^{-1}~\mathrm{s}$ 
(by adding $\Delta t_{0,B}\equiv t_{0,Sb2}-t_{0,Sb1}$ 
to the estimated $t_0$ for Sb2).
With this assumption, we find that using a value of $R_0 = 2.8\arcsec$ results in a 
model that agrees fairly well to shell Sb3.
The widths of the three shells, parameterized by $R_0$, are slightly different,
the fastest shell (Sb1) being the narrowest and the slowest shell (Sb3) being the widest.
Varying shell widths are needed for the model to fit the velocity gradient seen at the
edge of blue lobe cavity 
(e.g. at position offsets of about $6\arcsec$ on both sides of outflow axis) in the PV diagrams 
perpendicular to the outflow axis (panels b-e of Figure \ref{fig:pvdiagram}),
where the emission becomes wider at lower outflow velocities.

From the fitted models, it can be seen that the blue-shifted shells
in general are wider, slower and much older than the red-shifted shells.
On each side, the faster, younger shells are also narrower than the slower, older shells.
However, in the blue-shifted side, the shells have very similar widths 
(which can also be seen from the half opening angles listed in Table \ref{tab:model}),
while in the red-shifted side, the two shells have clearly different widths.
Furthermore, the three blue-shifted shells can be explained by outflow shells of
different dynamical ages, with similar age difference among consecutive shells.

\subsection{Origin of the Multiple Shell Structure}
\label{sec:multishell}

The parabolic outflowing shells can be produced by entrainment by a wide-angle wind
(\citealt[]{Li96,Lee00}).
In such models, the molecular outflow is swept up by a 
radial wide-angle wind with force distribution $\propto 1/\sin^2(\theta)$,
where $\theta$ is the polar angle relative to the outflow axis.
Such a wind interacts with a flattened ambient core with 
density distribution $\propto \sin^2(\theta)/r^2$,
and instantaneously mixes with shocked ambient gas. 
The resultant swept-up outflowing shell is then a radially expanding parabola with a 
Hubble law velocity structure.

Since each shell can be well fit with the wide-angle wind entrainment model,
it is natural to explain the multiple shell structure to be
formed by the entrainment of ambient circumstellar material by multiple outbursts of
a wide-angle wind.
One outburst of the wide-angle wind may not be able to 
entrain all the material to clear up the cavity, therefore
the later outbursts will continue to entrain material to form subsequent shells.
In such a scenario, the time intervals between successive shells,
which are $2.1\times 10^2~\mathrm{yr}$ for the red-shifted outflow and
$3.2\times 10^2~\mathrm{yr}$ for the blue-shifted outflow,
can be considered as the time interval between wind outbursts.
These estimated outburst intervals are consistent with those seen
in the episodic knots of HH 46/47 and in other sources. 
In the HH 46/47 outflow, an outburst interval of about 300 yr 
was estimated from the knots observed along the jet (see Paper I).
\citet[]{Plunkett15} estimated outburst intervals to range from 80 to 540 yr 
with a mean value of 310 yr  for a young embedded source in the 
Serpens South protostellar cluster. We thus suggest that in HH 46/47
the multiple shell structures may arise from the same high accretion rate episodes, which
is reflected in both the jet and wide-angle wind components of the outflow. 
In fact, in Paper I, the identified jet knots R1, R2 are found to have
dynamical ages of 360 and 650 yr, close to the ages of shells Sr1 (320 yr) and Sr2 (530 yr),
suggesting that the episodicity seen in the jet and the wide-angle outflows
may be caused by the same outburst events.

We note that the dynamical ages of these shells estimated here may not accurately reflect their true ages.
If the outburst happens in a short time compared to the dynamical time scale of the outflow shell, 
the shell entrained during
such an outburst event will decelerate due to the interaction with the surrounding material.
Therefore, it is likely that the estimated dynamical ages
are upper limits of their true ages.
The time intervals are also likely to be upper limits.
Yet, the similarity between the time intervals estimated here 
and those estimated from the jet supports the scenario that the different observed shell structures
are produced in multiple outburst events.

If the wide-angle shells on both sides are
caused by the same accretion bursts in the disk, 
then a shell in the blue-shifted side should correspond to
a shell in the red-shifted side.
However, it is difficult to identify such pairs in our data because the entrainment
is affected by significantly different environments with which each lobe is interacting. 
The dynamical ages of the identified shells in the blue-shifted lobe
are significantly larger than those of the shells in the red-shifted lobe and also significantly higher than the time interval
between shells ($\Delta t_{0,B}$). 
It is therefore likely that the Sb1, Sb2 or Sb3 shells on the blue-shifted lobe are 
not caused by the most recent outburst events. 
There may not be much molecular gas left in the blue lobe cavity in order for 
the youngest outburst to entrain any material to form a shell detectable in CO 
(unlike the red-shifted lobe, see below), as previous  
outflow bursts may have cleared the cavity.
Observations of the optical jet on the blue-shifted side found
that the furthermost jet knot (HH 47D) has a dynamical age of $1.3\times 10^3$ years (\citealt[]{Hartigan05}),
which is similar to the age of shell Sb1, and other knots closer to the protostar have much
younger ages. This supports that Sb1, Sb2 and Sb3 shells are not caused by 
recent, but by relatively old outburst events.
It is possible that the Sb4 and Sb5 structures in the blue-shifted side 
are caused by the most recent outburst, but due to the lack of ambient cloud  material,
the CO emission associated with these shell is only concentrated in the region close to the protostar.

Unlike the shells in the blue lobe, the youngest shell in the red lobe 
(Sr1) has an age similar to the outburst interval. Hence, the Sr1 shell may 
be the product of the most recent outburst. The red lobe is immersed in the 
dense part of the parent core and therefore there is still abundant material inside this cavity.
Also, the fact that Sr1 is significantly narrower than shell Sr2, 
is consistent with the scenario where Sr1 has only formed recently.
Since the narrower and newer shells are faster than the older and wider shells (see \S~4.1),
they are expected to collide with older shells in the future.
Based on the sizes $R_0$ and velocities ($v_0$; assumed to be constant),
Sb1 will catch up with Sb2 in $(R_{0,\mathrm{Sb2}}-R_{0,\mathrm{Sb1}})/
(v_{0,\mathrm{Sb1}}-v_{0,\mathrm{Sb2}})=2.7\times 10^2$ years, 
Sb2 will merge with Sb3 in $3.6\times 10^2$ years,
and Sr1 will reach Sr2 in $1.2\times 10^3$ years.
The real catch-up time scales should be shorter than these, as
the outer shells are likely to slow down due to the interaction with the dense ambient material.
This may explain why only two or three shells can be detected on both sides,
since the shells may only survive for a few outburst periods before
they collide with the old shells and form the outflow cavity walls seen
in the low-velocity channels.

In order to further explore whether the observed outflow shells are 
caused by entrainment/interaction with the envelope
or are being directly launched from the disk,
we estimate the mass and momentum rates of these shells from the $\twcob$ emission.
To obtain the gas mass, we assume optically thin emission and adopt an
abundance of $^{12}$CO of $10^{-4}$ relative to H$_2$ and a gas mass
of $2.34\times10^{-24}$ g per H$_2$ molecule.  
Following Paper II, we adopt an
excitation temperature of $\tex=15~\K$.  An excitation temperature of
50~K would increase the mass estimate by a factor of 1.5.  
In each velocity channel, we only include the primary beam corrected emission 
above $3\sigma$ and within a primary beam response greater than $0.2$
relative to the phase center.
We include all the emission associated with the outflow,
except the emission at outflow velocities less than
$2~\kms$ in order to 
avoid possibly adding emission from core material to our outflow mass estimate.

We estimate a total mass of  $5.6\times 10^{-3}$ and $1.0\times 10^{-2}~M_\odot$
and momentum of $4.4\times 10^{-2}$ and $8.7\times 10^{-2}~M_\odot~\kms$
in the shells of the blue- and red-shifted outflow lobes, respectively.
Here, in calculating the total momenta, 
we use the velocity of each channel and multiply by the outflow mass of that channel.
These are very likely lower limits due to the optically thin assumption,
uncounted low-velocity outflowing material
and possible higher excitation temperatures (e.g., \citealt[]{Dunham14}).
Using the estimated ages of the oldest shells on both sides ($1.8\times10^3$ yr for 
Sb3 and $5.3\times10^2$ yr for Sr2),
the time-averaged mass outflow rates are $3.1\times10^{-6}$
and $1.8\times10^{-5}~M_\odot~\yr^{-1}$
for the blue- and red-shifted outflow lobes, respectively.
And the time-averaged momentum injection rates are $2.4\times10^{-5}$
and $1.6\times10^{-4}~M_\odot~\yr^{-1}~\kms$
for the blue and red-shifted outflow lobes, respectively.
Note that these rates are averaged over the outflow age, and the mass loss and momentum
injection rates during each outburst are expected to be significantly higher than these values.

The above estimates for the mass outflow rates are one to two orders of magnitude larger 
(or even larger given that the values quoted above are lower limits) than 
most estimates of the mass loss rate for the HH 46/47 protostellar jet using optical and 
IR atomic line emission, which range between 0.3 and $5 \times 10^{-7}~M_\odot~\yr^{-1}$
(e.g. \citealt[]{Hartigan94,Antoniucci08,Garcia10,Mottram17})
\footnote{These values are consistent with the ``typical'' mass-loss rate value for winds in 
Class I sources of $\sim 10^{-7}~M_\odot~\yr^{-1}$ 
(e.g. \citealt[]{Hartigan94,Podio06,Podio12,Mottram17}. 
The mass loss rate estimate for the HH 46/47 jet of $(2-9) \times 10^{-6}~M_\odot~\yr^{-1}$ 
quoted by \citet[]{Nisini15} is an outlier compared to other measurements in the literature.}.
Moreover, if we assume a mass loss rate of 
$\sim 10^{-7}~M_\odot~\yr^{-1}$ and a velocity of about  $100~\kms$ 
(e.g. \citealt[]{Morse94,Hartigan05}) for the wind launched by the disk,
this then leads to a  momentum injection rate of approximately $10^{-5}~M_\odot~\yr^{-1}$.
These results show that the observed $\twcob$ shells have mass loading rates
that are one to two orders of magnitudes higher than the mass loss rates of the 
jet (or wind) launched from the disk, but
have momentum injection rates similar to the jet/winds directly launched from
the disk. This is consistent with the  scenario in which the observed 
$\twcob$ shells 
are mostly made of
ambient material that was
entrained by the wind launched from the disk,
in a momentum-conserving interaction, and 
not of material that was directly
launched from the disk.
This is also consistent with theoretical simulations which show that only $25\%-30\%$ of the mass
in a molecular outflow is  directly launched from the disk and the rest of the mass is entrained material (e.g. \citealt[]{Offner17})

As discussed above, the observed shells are consistent with wide-angle wind
entrainment with multiple outburst events. Although the same episodicity is also seen in
the jet, we think these shells are unlikely to be formed by jet entrainment.
In fact, multiple layer structures were identified in the extended red-shifted lobe 
of the HH 46/47 molecular outflow observed in $\twcoa$,
which were identified to be associated with several jet bow-shock events (see Paper II).
Those structures are at much lower velocities ($\vout < 10~\kms$), have a different 
morphology and are found at much larger distances from the source ($> 50\arcsec$) 
compared to the shells reported here. Thus they are unlikely to be associated to 
the high-velocity shells discussed in this paper.
In some cases, shells at the base of the outflow cavities indeed are found
to be connected to the jet bow-shocks far away from the central sources (e.g. \citealt[]{Lee15}).
However, it is unclear 
whether the morphology and kinematics of the shells observed here 
(which are consistent with those expected for 
radially expanding parabolic shells), can be also explained by jet bow-shock entrainment. 
More theoretical simulations and models are needed to test whether
such shells can be formed solely by jet bow-shock entrainment.

\subsection{Implications for Evolution of Protostellar Outflow}

The opening angle of protostellar outflows appears to increase
with the source's age;  the outflow cavity gradually widens as the source evolves
(e.g. \citealt[]{Arce06,Seale08,Velusamy14,Hsieh17}).
Outflows are therefore thought to be able to  disperse the parent core,
terminate the accretion phase, and
regulate the core-to-star formation efficiency (e.g. \citealt[]{Machida13,Offner14,Offner17}).
There are several ways that an outflow can widen as it evolves.
In one scenario the outflow cavity widens as the envelope material is 
continuously entrained by the protostellar jet and/or wide-angle disk wind.
In this model one would expect the 
the recently accelerated
material to be inside of the older, previously entrained, material.
The newest and faster shell will soon reach the outer, slower, shells and 
transfer momentum to them and the outflow cavity walls. 
This way, in general, the outflow cavity opening angle would be increasing with time.
The observed multiple shell structure in HH 46/47 appears to be consistent with this picture. 

In the second scenario the observed outflow is mostly composed of 
material that is directly  launched from the disk (e.g. \citealt[]{Machida13}). 
In this model, the disk slowly grows in size, and with it the launching region, 
at the base of the outflow, slowly widens. This in turn produces the outflow cavity to 
gradually become less collimated. 
In this scenario at least a part of the recently launched material is expected to be outside of the 
previously launched material, which is launched from the new outer regions of the disk.
Such model, however, is not consistent with the observations presented here,
in which the molecular outflow is made of entrained material and the material 
entrained by the most recent outflow episode is inside of the older shells.
However, we note that these two scenarios are not mutually exclusive.

It is also possible that the outflow cavity widens as the outflow changes direction 
over time (e.g. \citealt[]{Offner16,LeeJ17}), which can be caused by a change in the angular momentum direction
of the accretion flow, binary interaction, and/or jet precession. 
However, in the case of HH 46/47, despite  the existence of a binary system at the center,
the main outflow appears to be symmetric and not affected by a secondary outflow (Papers I and II).
Also the precession of the jet appears to be much smaller than the opening angle of the outflow
cavity (Paper I), which also indicates that changing outflow direction may not be the dominant
cause for the widening of the outflow in HH 46/47.

\subsection{Implications for Wind Launching}

The spatial resolution of our current observation is too low to resolve the 
launching region of the wide-angle wind that entrained the observed molecular outflow.
However, the highly coherent properties of the observed outflow shells can provide some constraints
on the wind launching mechanism.
In the outflow entrainment scenario, in order to form such a coherent shell structure in
each outburst, the wind launched from the disk towards different polar angles needs to be well coordinated.
Such coordination can be naturally understood if the launching area in each outburst 
is a narrow region on the disk.
It is very likely that the duration of each outburst is significantly shorter than the 
interval between outburst event, given that the observed outflow shell from each outburst is very well defined.
We, therefore, assume the duration of each outburst $\Delta t_\mathrm{outburst}$ is $\sim 20\%-30\%$  
of the outburst intervals or about 60 years (see \S\ref{sec:fitting}).
If we use the sound speed $c_s$ as the characteristic speed of accreting material in the disk moving
inward, we obtain a length scale of 
$\Delta R=c_s\Delta t_\mathrm{outburst}=(6~\au)\sqrt{T/40~\K}$,
which we can use as a proxy for the width of the launching region on the disk.
If the width of the outflow launching region is much larger than this, it would then be hard
for the wind launching at different stellocentric radii to be well coordinated to form such coherent shells.
This estimate of the size for the outflow launching region is consistent with  recent observational studies that deduce a relatively narrow range of radii for the outflow launching regions
(e.g. \citealt[]{Bjerkeli16,Hirota17,Zhang18}).
Note that this is different from the classical picture of a disk wind
that is launched over a wide range of stellocentric radii (e.g. \citealt[]{BP82}).

Since accretion variability is believed to be caused by various instabilities in the accretion disk,
it is possible that such instabilities can affect particular regions on the disk to enhance the
mass or momentum of the launched wind during the outburst.
We can further use the outburst interval of $200-300$ years to obtain a characteristic 
radius by comparing the outburst interval with the Keplerian orbital timescales. 
The resultant radius is $10-13$ au,
assuming a mass of $0.3~M_\odot$ for the central object (see Paper II).
This can be used as a characteristic radius for the disk instability and outflow launching zone.
Note that this source contains a proto-binary system with a separation of 0.26$\arcsec$ (120 au)
at the center, therefore the estimated characteristic radius of $10-13$ au indicates that
the outflow launching region is likely on a circumstellar disk around the primary.

\section{Conclusions}
\label{sec:conclusion}

We present ALMA $\twcob$ observation of the HH 46/47 molecular outflow, 
in which we have detected multiple wide-angle outflowing shell structures 
in both the blue and red-shifted lobes. 
These shells are found to be highly coherent in  position-position-velocity space, 
extending to $\gtrsim 40-50~\kms$ in velocity and $10^4~\au$ in space 
with well defined morphology and kinematics.
We argue that these structures are formed by the entrainment of circumstellar gas by a wide-angle wind with multiple outburst events.
The intervals between consecutive  outburst are found to be $(2-3)\times 10^2~\mathrm{yr}$, 
consistent with the timescale between outburst events  in the jet powered by the same protostar.
Our results provide strong evidence that  wide-angle disk
winds can be episodic, just as protostellar jets.

\acknowledgements 
The authors thank Nami Sakai and Satoshi Yamamoto for valuable discussions.
This paper makes use of the following ALMA data: ADS/JAO.ALMA\#2015.1.01068.S. 
ALMA is a partnership of ESO (representing its member states), NSF (USA) and NINS (Japan), 
together with NRC (Canada) and NSC and ASIAA (Taiwan) and KASI (Republic of Korea), 
in cooperation with the Republic of Chile. 
The Joint ALMA Observatory is operated by ESO, AUI/NRAO and NAOJ.
The National Radio Astronomy Observatory is a facility of the National Science Foundation 
operated under cooperative agreement by Associated Universities, Inc.
Y.Z. acknowledges support from RIKEN Special Postdoctoral Researcher Program.
H.G.A acknowledges support from his NSF grant AST-1714710.
D.M. and G.G. acknowledge support from CONICYT project Basal AFB-170002.

\software{CASA (\citealt[]{McMullin07})}

\end{document}